\documentclass{elsart}
\usepackage{amsmath}
\newcommand{\Li}[1]{\mathop{\mathrm{Li}}\nolimits_{#1}}

\begin{document}

\begin{frontmatter}
\title{Matching heavy-quark fields in QCD and HQET at three loops}
\author{A.G.~Grozin}
\address{Institut f\"ur Theoretische Teilchenphysik,
Karlsruher Institut f\"{u}r Technologie, 76128 Karlsruhe, Germany}

\begin{abstract}
The relation between the heavy-quark field in QCD
and the corresponding field in HQET
is derived up to three loops,
and to all orders in the large-$\beta_0$ limit.
The corresponding relation between the QED electron field
and the Bloch--Nordsieck one is gauge invariant to all orders.
We also prove that the $\overline{\text{MS}}$ anomalous dimension
of the QED electron field depends on the gauge parameter only at one loop.
\end{abstract}
\end{frontmatter}

QCD problems with a single heavy quark $Q$
can be treated in a simpler effective theory --- HQET,
if there exists a 4-velocity $v$ such that
the heavy-quark momentum is $p=mv+k$
($m$ is the on-shell mass)
and the characteristic residual momentum is small: $k\ll m$.
QCD operators can be written as series in $1/m$ via HQET operators;
the coefficients in these series are determined by matching
on-shell matrix elements in both theories.
For example, the heavy--light quark currents have been considered
at the leading (zeroth) order in $1/m$
up to three-loop accuracy~\cite{BG:95,BGMPSS:10}.

Here we shall consider the heavy-quark field $Q$.
Though its matrix elements are not directly observable,
its matching coefficient can have some applications.
For example, it is not possible to simulate heavy quarks on a lattice directly,
because at present we cannot have lattice spacings $a\ll1/m$.
On the other hand, simulating HQET on a lattice
only requires $a\ll1/\Lambda_{\overline{\text{MS}}}$.
It is possible to obtain the HQET heavy-quark propagator
in the Landau gauge from such simulations.
Then, if we know the matching coefficient,
we can reconstruct a fundamental QCD quantity ---
the heavy-quark propagator as a function of $x$.

\begin{sloppypar}
At the tree level, $Q$ is related to the corresponding HQET field $Q_v$
(satisfying $\rlap/v Q_v=Q_v$) by~\cite{Lee:91,KT:91,MRR:92,N:94}
\begin{equation}
Q(x) = e^{-i m v\cdot x}
\left(1 + \frac{i \rlap{\hspace{0.2em}/}D_\bot}{2m} + \cdots\right)
Q_v(x)\,,\quad
D^\mu_\bot=D^\mu-v^\mu v\cdot D\,.
\label{tree}
\end{equation}
The matrix elements of the bare fields
between the on-shell quark with momentum $p=mv+k$ and the vacuum in both theories
are given by the on-shell wave-function renormalization constants:
\begin{equation}
{<}0|Q_0|Q(p){>} = \left(Z_Q^{\text{os}}\right)^{1/2} u(p)\,,\quad
{<}0|Q_{v0}|Q(p){>} = \left(\tilde{Z}_Q^{\text{os}}\right)^{1/2} u_v(k)
\label{onshell}
\end{equation}
(HQET renormalization constants are denoted by $\tilde{Z}$).
The Dirac spinors are related by the Foldy--Wouthuysen transformation
\begin{equation*}
u(mv+k) = \left[1 + \frac{\rlap/k}{2m}
+ \mathcal{O}\left(\frac{k^2}{m^2}\right) \right] u_v(k)\,.
\end{equation*}
Therefore, the bare fields are related by
\begin{equation}
Q_0(x) = e^{-i m v\cdot x} \left[ z_0^{1/2}
\left(1 + \frac{i \rlap{\hspace{0.2em}/}D_\bot}{2m}\right) Q_{v0}(x)
+ \mathcal{O}\left(\frac{1}{m^2}\right) \right]\,,
\label{bare}
\end{equation}
where the bare matching coefficient is
\begin{equation}
z_0 = \frac{Z_Q^{\text{os}}(g_0^{(n_l+1)},a_0^{(n_l+1)})}%
{\tilde{Z}_Q^{\text{os}}(g_0^{(n_l)},a_0^{(n_l)})}
\label{z0}
\end{equation}
(we use the covariant gauge: the gauge-fixing term in the Lagrangian
is $-(\partial_\mu A_0^{a\mu})/(2a_0)$, and the free gluon propagator
is $(-i/p^2) (g_{\mu\nu} - (1-a_0) p_\mu p_\nu/p^2)$;
the number of flavours in QCD is $n_f=n_l+1$).
The $\mathcal{O}(1/m)$ matching coefficient in~(\ref{bare})
is equal to the leading one, $z_0$;
this reflexes the reparametrization invariance~\cite{LM:92}.
The $\overline{\text{MS}}$ renormalized fields are related
by the formula similar to~(\ref{bare}),
with the renormalized decoupling coefficient
\begin{equation}
z(\mu) =
\frac{\tilde{Z}_Q(\alpha_s^{(n_l)}(\mu),a^{(n_l)}(\mu))}%
{Z_Q(\alpha_s^{(n_l+1)}(\mu),a^{(n_l+1)}(\mu))} z_0\,.
\label{renorm}
\end{equation}
\end{sloppypar}

If there are no massive flavours except $Q$,
then $\tilde{Z}_Q^{\text{os}}=1$
because all loop corrections are scale-free.
The QCD on-shell renormalization constant $Z_Q^{\text{os}}$
contains the single scale $m$ in this case;
it has been calculated~\cite{MR:00} up to three loops.
The three-loop $\overline{\text{MS}}$ anomalous dimensions
of $Q_v$~\cite{MR:00,CG:03} and $Q$~\cite{T:82,LV:93} are also known.
We have to express all three quantities
$Z_Q^{\text{os}}(g_0^{(n_l+1)},a_0^{(n_l+1)})$,
$Z_Q(\alpha_s^{(n_l+1)}(\mu),a^{(n_l+1)}(\mu))$,
$\tilde{Z}_Q(\alpha_s^{(n_l)}(\mu),a^{(n_l)}(\mu))$
via the same variables, say,
$\alpha_s^{(n_l)}(\mu)$, $a^{(n_l)}(\mu)$,
see~\cite{CKS:98}.
An explicit formula expressing $\alpha_s^{(n_l+1)}(\mu)$
via $\alpha_s^{(n_l)}(\mu)$ and $L = 2 \log(\mu/m)$
($m$ is the on-shell mass) can be found in~\cite{GMPS:08}.
The corresponding relation between $a^{(n_l+1)}(\mu)$
and $a^{(n_l)}(\mu)$ is
\begin{equation}
\begin{split}
&\frac{a^{(n_l+1)}(\mu)}{a^{(n_l)}(\mu)} = 1
- \biggl[ \frac{4}{3} L
    + \frac{6 L^2 + \pi^2}{9} \varepsilon
    + \frac{2 L^3 + \pi^2 L - 4 \zeta_3}{9} \varepsilon^2
    + \cdots \biggr]
T_F \frac{\alpha_s^{(n_l)}(\mu)}{4\pi}\\
&{} - \biggl[ C_A L^2 + \left(4 C_F + 5 C_A\right) L
+ 15 C_F - \frac{13}{12} C_A\\
&\hphantom{{}-\biggl[\biggr.}{} +
\biggl( C_A L^3 + \left(4 C_F + 5 C_A\right) L^2
    + \biggl(30 C_F + \frac{\pi^2-13}{6} C_A\biggr) L\\
&\hphantom{{}-\biggl[{}+\biggl(\biggr.\biggr.}{}
    + \biggl(\frac{\pi^2}{3} + \frac{31}{2}\biggr) C_F
    + \frac{5 \pi^2 + 169}{12} C_A
\biggr) \varepsilon + \cdots \biggr]
T_F \left(\frac{\alpha_s^{(n_l)}(\mu)}{4\pi}\right)^2
+ \cdots
\end{split}
\label{dec}
\end{equation}

Our main result is the renormalized matching coefficient
\begin{equation}
\begin{split}
z(\mu) =& 1 - (3 L + 4) C_F \frac{\alpha_s^{(n_l)}(\mu)}{4\pi}
+ \left(z_{22} L^2 + z_{21} L + z_{20}\right)
C_F \left(\frac{\alpha_s^{(n_l)}(\mu)}{4\pi}\right)^2\\
&{} + \left(z_{33} L^3 + z_{32} L^2 + z_{31} L + z_{30}\right)
C_F \left(\frac{\alpha_s^{(n_l)}(\mu)}{4\pi}\right)^3 + \cdots
\end{split}
\label{zmu}
\end{equation}
where
\begin{equation*}
\begin{split}
z_{22} =& \frac{9}{2} C_F - \frac{11}{2} C_A + 2 T_F n_l\,,\\
z_{21} =& \frac{27}{2} C_F - \frac{215}{6} C_A
+ \frac{38}{3} T_F n_l + 2 T_F\,,\\
z_{20} =&
\left(16 \pi^2 \log2 - 24 \zeta_3 - 13 \pi^2 + \frac{241}{8}\right) C_F\\
&{}+ \left(- 8 \pi^2 \log2 + 12 \zeta_3 + 5 \pi^2 - \frac{1705}{24}\right) C_A\\
&{} + \left(\frac{4}{3} \pi^2 + \frac{113}{6}\right) T_F n_l
+ \left(- \frac{16}{3} \pi^2 + \frac{947}{18}\right) T_F\,,\\
z_{33} =& - \frac{9}{2} C_F^2 + \frac{33}{2} C_F C_A - \frac{121}{9} C_A^2
- 6 C_F T_F n_l\\
&{} + \frac{88}{9} C_A T_F n_l + \frac{1}{3} a^{(n_l)}(\mu) C_A T_F
- \frac{16}{9} T_F^2 n_l^2\,,\\
z_{32} =& - \frac{45}{2} C_F^2 + 135 C_F C_A - \frac{2671}{18} C_A^2
- 42 C_F T_F n_l - 4 C_F T_F\\
&{} + \frac{938}{9} C_A T_F n_l
- \biggl(\frac{13}{6} a^{(n_l)}(\mu) + 1\biggr) C_A T_F
- \frac{152}{9} T_F^2 n_l^2 + \frac{8}{3} T_F^2\,,\\
z_{31} =&
\biggl[- 48 \pi^2 \log2 + 72 \zeta_3 + 39 \pi^2 - \frac{783}{8}\biggr] C_F^2\\
&{} + \biggl[\frac{424}{3} \pi^2 \log2 - 224 \zeta_3 - \frac{331}{3} \pi^2
    + \frac{13307}{24}\biggr] C_F C_A
\end{split}
\end{equation*}
\begin{equation*}
\begin{split}
&{} + \biggl[
\biggl(- \frac{2}{45} \pi^4 + \frac{9}{4} \zeta_3 + \frac{1}{4} \biggr)
a^{(n_l)}(\mu)\\
&\hphantom{{}+\biggl[\biggr.}{}
- \frac{176}{3} \pi^2 \log2 + \frac{325}{4} \zeta_3
- \frac{2}{15} \pi^4 + \frac{110}{3} \pi^2 - \frac{73981}{108}
\biggr] C_A^2\\
&{} + \biggl( - \frac{128}{3} \pi^2 \log2 + 16 \zeta_3 + \frac{92}{3} \pi^2
- \frac{613}{6}\biggr) C_F T_F n_l
+ \biggl(16 \pi^2 - \frac{1013}{6}\biggr) C_F T_F\\
&{} + \biggl(\frac{64}{3} \pi^2 \log2 + 16 \zeta_3 - \frac{32}{9} \pi^2
+ \frac{10816}{27}\biggr) C_A T_F n_l\\
&{} + \biggl(\frac{121}{18} a^{(n_l)}(\mu) - \frac{352}{9} \pi^2
+ \frac{11278}{27}\biggr) C_A T_F\\
&{}
- \biggl(\frac{32}{9} \pi^2 + \frac{1336}{27}\biggr) T_F^2 n_l^2
+ \biggl(\frac{128}{9} \pi^2 - \frac{3908}{27}\biggr) T_F^2 n_l
- \frac{20}{9} T_F^2\,,\\
z_{30} =& \biggl[ - 1792 a_4 - \frac{224}{3} \log^4 2
+ 96 \pi^2 \log^2 2 + \frac{3568}{3} \pi^2 \log2
- 20 \zeta_5\\
&\hphantom{{}+\Biggl\{\Biggr.}{}
+ 8 \pi^2 \zeta_3 - 1256 \zeta_3
- \frac{76}{15} \pi^4 - \frac{4801}{9} \pi^2
- \frac{3023}{12} \biggr] C_F^2\\
&{} + \biggl[ - \frac{32}{3} a_4 - \frac{4}{9} \log^4 2
- \frac{1448}{9} \pi^2 \log^2 2 - \frac{2752}{9} \pi^2 \log2
+ 580 \zeta_5\\
&\hphantom{{}+\Biggl\{\Biggr.}{}
- 180 \pi^2 \zeta_3 - \frac{2312}{3} \zeta_3
+ \frac{6697}{270} \pi^4 + \frac{2137}{9} \pi^2
+ \frac{24131}{72} \biggr] C_F C_A\\
&{} + \biggl[ \biggl( - \frac{7}{6} \zeta_5 - \frac{4}{9} \pi^2 \zeta_3
+ \frac{13}{4} \zeta_3 - \frac{17}{432} \pi^4 + \frac{1}{4} \pi^2
+ \frac{13}{12}\biggr) a^{(n_l)}(\mu)\\
&\hphantom{{}+\Biggl\{\Biggr.}{}
+ \frac{1360}{3} a_4 + \frac{170}{9} \log^4 2
+ \frac{508}{9} \pi^2 \log^2 2 - \frac{1300}{9} \pi^2 \log2
- \frac{787}{2} \zeta_5\\
&\hphantom{{}+\Biggl\{\Biggr.}{}
+ \frac{340}{3} \pi^2 \zeta_3 + \frac{23311}{36} \zeta_3
- \frac{20429}{2160} \pi^4 - \frac{8705}{108} \pi^2 - \frac{1656817}{1944}
\biggr] C_A^2\\
&{} + \biggl[ \frac{1024}{3} a_4 + \frac{128}{9} \log^4 2
+ \frac{256}{9} \pi^2 \log^2 2 - \frac{1504}{9} \pi^2 \log2\\
&\hphantom{{}+\Biggl\{\Biggr.}{}
+ \frac{1096}{3} \zeta_3 - \frac{916}{135} \pi^4
+ \frac{904}{9} \pi^2 + \frac{1120}{9}
\biggr] C_F T_F n_l\\
&{} + \biggl[ 768 a_4 + 32 \log^4 2
- 32 \pi^2 \log^2 2 + \frac{1088}{9} \pi^2 \log2\\
&\hphantom{{}+\Biggl\{\Biggr.}{}
+ \frac{466}{9} \zeta_3 + \frac{124}{45} \pi^4
- \frac{8848}{81} \pi^2 - \frac{16811}{54}
\biggr] C_F T_F\\
&{} + \biggl[ - \frac{512}{3} a_4 - \frac{64}{9} \log^4 2
- \frac{128}{9} \pi^2 \log^2 2 + \frac{752}{9} \pi^2 \log2\\
&\hphantom{{}+\Biggl\{\Biggr.}{}
- \frac{280}{9} \zeta_3 + \frac{152}{135} \pi^4
+ \frac{52}{3} \pi^2 + \frac{111791}{243}
\biggr] C_A T_F n_l
\end{split}
\end{equation*}
\begin{equation*}
\begin{split}
&{} + \biggl[ \biggl(\frac{8}{3} \zeta_3 - \frac{2461}{108}\biggr) a^{(n_l)}(\mu)
- 512 a_4 - \frac{64}{3} \log^4 2
+ \frac{64}{3} \pi^2 \log^2 2 + \frac{5120}{9} \pi^2 \log2\\
&\hphantom{{}+\Biggl\{\Biggr.}{}
- 60 \zeta_5 + \frac{44}{3} \pi^2 \zeta_3 - \frac{2837}{9} \zeta_3
- \frac{136}{45} \pi^4 - \frac{36268}{81} \pi^2 + \frac{100627}{81}
\biggr] C_A T_F\\
&{} - \biggl(\frac{224}{9} \zeta_3 + \frac{304}{27} \pi^2
+ \frac{11534}{243}\biggr) T_F^2 n_l^2
+ \biggl(\frac{208}{9} \pi^2 - \frac{18884}{81}\biggr) T_F^2 n_l\\
&{} + \biggl(\frac{448}{3} \zeta_3 + \frac{128}{45} \pi^2
- \frac{16850}{81}\biggr) T_F^2
\end{split}
\end{equation*}
(here $a_4=\Li4(1/2)$).
Gauge dependence first appears at three loops,
as in $Z_Q^{\text{os}}$~\cite{MR:00}.
The requirement of finiteness
of the renormalized matching coefficient~(\ref{renorm})
at $\varepsilon\to0$ has allowed the authors of~\cite{MR:00}
to extract $\tilde{Z}_Q$ from their result for $Z_Q^{\text{os}}$.

It would not be too difficult to take into account
a lighter massive flavour, say, $m_c\neq0$ in $b$-quark HQET.
$\tilde{Z}_Q^{\text{os}}$ is no longer equal to 1,
but is known at three loops~\cite{GSS:06};
$Z_Q^{\text{os}}$ contains two scales,
and is a non-trivial function of $m_c/m_b$~\cite{BGSS:07}.
Both $\tilde{Z}_Q^{\text{os}}$ and $Z_Q^{\text{os}}$
have no smooth limit at $m_c\to0$,
but the discontinuity cancels in the ratio~(\ref{z0}).

Now let's consider $z(\mu)$ in the large-$\beta_0$ limit
(see Chapter~8 in~\cite{G:04} for a pedagogical introduction):
\begin{equation}
z(\mu) = 1 + \int_0^\beta \frac{d\beta}{\beta}
\left(\frac{\gamma(\beta)}{2\beta} - \frac{\gamma_0}{2\beta_0}\right)
+ \frac{1}{\beta_0} \int_0^\infty du\,e^{-u/\beta} S(u)
+ \mathcal{O}\left(\frac{1}{\beta_0^2}\right)\,,
\label{largeb0}
\end{equation}
where $\beta=\beta_0\alpha_s/(4\pi)$,
$\gamma=\gamma_0\alpha_s/(4\pi)+\cdots$
(differences of $n_l$-flavour and $(n_l+1)$-flavour quantities
can be neglected at the $1/\beta_0$ order).
The difference of the QCD and HQET anomalous dimensions
$\gamma=\gamma_Q-\tilde{\gamma}_Q$
and the Borel image $S(u)$ can be expressed as
\begin{equation}
\gamma(\beta) = - 2 \frac{\beta}{\beta_0} F(-\beta,0)\,,\quad
S(u) = \frac{F(0,u)-F(0,0)}{u}\,,
\label{struct}
\end{equation}
where the function $F(\varepsilon,u)$ has been calculated in~\cite{BG:95}
(see also~\cite{G:04}):
\begin{equation}
\begin{split}
F(\varepsilon,u) =& - 2 C_F
\left(\frac{\mu}{m}\right)^{2u} e^{\gamma_E\varepsilon}
\frac{\Gamma(1+u) \Gamma(1-2u)}{\Gamma(3-u-\varepsilon)}
D(\varepsilon)^{u/\varepsilon-1}\\
&{}\times(3-2\varepsilon) (1-u) (1+u-\varepsilon)\,,\\
D(\varepsilon) ={}& 6 e^{\gamma_E\varepsilon}
\Gamma(1+\varepsilon) B(2-\varepsilon,2-\varepsilon)
= 1 + \frac{5}{3} \varepsilon + \cdots
\end{split}
\label{Feu}
\end{equation}
The anomalous dimension difference~\cite{BG:95} is gauge invariant
at this order:
\begin{equation}
\gamma(\beta) = 2 C_F \frac{\beta}{\beta_0}
\frac{(1+\beta) (1+\frac{2}{3}\beta)}%
{B(2+\beta,2+\beta) \Gamma(3+\beta) \Gamma(1-\beta)}\,;
\label{gamma}
\end{equation}
the Borel image is~\cite{NS:95,G:04}
\begin{equation}
S(u) = - 6 C_F \left[ e^{(L+5/3)u}
\frac{\Gamma(u) \Gamma(1-2u)}{\Gamma(3-u)} (1-u^2)
- \frac{1}{2u} \right]\,.
\label{Su}
\end{equation}

This Borel image has infrared renormalon poles at each positive
half-integer $u$ and at $u=2$.
Therefore, the integral in~(\ref{largeb0}) is not well defined.
Comparing its residue at the leading pole $u=1/2$
with the residue of the static-quark self-energy at its
ultraviolet pole $u=1/2$~\cite{BB:94},
we can express the renormalon ambiguity of $z(\mu)$ as
\begin{equation}
\Delta z(\mu) = \frac{3}{2} \frac{\Delta\bar{\Lambda}}{m}
\label{ambig}
\end{equation}
($\bar{\Lambda}$ is the ground-state meson residual energy).
This ambiguity is compensated in physical matrix elements
by ultraviolet renormalon ambiguities in the leading $1/m$ correction
(matrix elements of both local and bilocal dimension-5/2 operators),
see~\cite{NS:95}.

The matching coefficient is gauge invariant at the order $1/\beta_0$.
Expanding $\gamma(\beta)$ and $S(u)$ and integrating,
we obtain
\begin{equation}
\begin{split}
&z(\mu) = 1 - C_F \frac{\alpha_s(\mu)}{4\pi} \Biggl\{ 3 L + 4
+ \biggl[ \frac{3}{2} L^2 + \frac{19}{2} L + \pi^2 + \frac{113}{8} \biggr]
\frac{\beta_0 \alpha_s(\mu)}{4\pi}\\
&{} + \biggl[ L^3 + \frac{19}{2} L^2
+ \biggl( 2 \pi^2 + \frac{167}{6} \biggr) L
+ 14 \zeta_3 + \frac{19}{3} \pi^2 + \frac{5767}{216} \biggr]
\left(\frac{\beta_0 \alpha_s(\mu)}{4\pi}\right)^2\\
&{} + \biggl[ \frac{3}{4} L^4 + \frac{19}{2} L^3
+ \biggl( 3 \pi^2 + \frac{167}{4} \biggr) L^2
+ \biggl( 36 \zeta_3 + 19 \pi^2 + \frac{2903}{36} \biggr) L\\
&\hphantom{{}+\biggl[\biggr.}{}
+ \frac{71}{40} \pi^4 + \frac{467}{4} \zeta_3 + \frac{167}{6} \pi^2
+ \frac{103933}{1728} \biggr]
\left(\frac{\beta_0 \alpha_s(\mu)}{4\pi}\right)^3
+ \cdots \Biggr\}\,.
\end{split}
\label{beta}
\end{equation}
Thus we have confirmed the contributions with the highest power of $n_l$
in each term in~(\ref{zmu}),
and predicted such a contribution at $\alpha_s^4$.

Numerically, the matching coefficient~(\ref{zmu})
in the Landau gauge at $\mu=m$ and $n_l=4$ can be written as
\begin{equation*}
\begin{split}
z(m) ={}& 1 - \frac{4}{3} \frac{\alpha_s^{(4)}(m)}{\pi}
- (1.9996 \beta_0 - 4.5421)
\left(\frac{\alpha_s^{(4)}(m)}{\pi}\right)^2\\
&{} - (2.2091 \beta_0^2 + 5.1153 \beta_0 - 61.5397)
\left(\frac{\alpha_s^{(4)}(m)}{\pi}\right)^3\\
&{} - (3.3755 \beta_0^3 + \cdots)
\left(\frac{\alpha_s^{(4)}(m)}{\pi}\right)^4
+ \cdots
\end{split}
\end{equation*}
\begin{equation}
\begin{split}
{}={}& 1 - \frac{4}{3} \frac{\alpha_s^{(4)}(m)}{\pi}
- (16.6629 - 4.5421)
\left(\frac{\alpha_s^{(4)}(m)}{\pi}\right)^2\\
&{} - (153.4076 + 42.6271 - 61.5397)
\left(\frac{\alpha_s^{(4)}(m)}{\pi}\right)^3\\
&{} - (1953.4013 + \cdots)
\left(\frac{\alpha_s^{(4)}(m)}{\pi}\right)^4
+ \cdots\\
{}={}& 1 - \frac{4}{3} \frac{\alpha_s^{(4)}(m)}{\pi}
- 12.1208
\left(\frac{\alpha_s^{(4)}(m)}{\pi}\right)^2
- 134.4950
\left(\frac{\alpha_s^{(4)}(m)}{\pi}\right)^3\\
&{} - (1953.4013 + \cdots)
\left(\frac{\alpha_s^{(4)}(m)}{\pi}\right)^4
+ \cdots
\end{split}
\label{numeric}
\end{equation}
($\beta_0$ is for $n_l=4$ flavours).
Naive nonabelianization~\cite{BG:95} works rather well at two
and three loops
(in the latter case the $\mathcal{O}(\beta_0)$ and $\mathcal{O}(1)$
terms partially compensate each other, similarly to~\cite{BGMPSS:10}).
Therefore, we can expect that the estimate of the $\alpha_s^4$
term is also reasonably good.
Numerical convergence of the series is very poor;
this is related to the infrared renormalon at $u=1/2$.

Now let us consider the relation between the $\overline{\text{MS}}$ renormalized
electron field in QED and the  Bloch--Nordsieck electron field~\cite{BN:37}.
The bare matching coefficient $z_0=Z_\psi^{\text{os}}$
is gauge invariant to all orders~\cite{LK:56,JZ:59,MR:00}.
In the Bloch-Nordsieck model, due to exponentiation~\cite{YFS:61}, 
$\log\tilde{Z}_\psi=(3-a^{(0)})\alpha^{(0)}/(4\pi\varepsilon)$
(where the 0-flavour $\alpha^{(0)}$ is equal to the on-shell $\alpha\approx1/137$).
In the full QED, $\log Z_\psi=-a^{(1)}\alpha^{(1)}/(4\pi\varepsilon)
+(\text{gauge-invariant higher terms})$, see the Appendix.
The gauge dependence cancels in $\log(\tilde{Z}_\psi/Z_\psi)$
because of the QED decoupling relation $a^{(1)}\alpha^{(1)}=a^{(0)}\alpha^{(0)}$.
Therefore, the renormalized matching coefficient $z(\mu)$ in QED
is gauge invariant to all orders.
We obtain
\begin{equation}
\begin{split}
z(\mu) =& 1 - (3 L + 4) \frac{\alpha}{4\pi}\\
&{} + \biggl(\frac{9}{2} L^2 + \frac{31}{2} L
+ 16 \pi^2 \log2 - 24 \zeta_3 - \frac{55}{3} \pi^2 + \frac{5957}{72}\biggr)
\left(\frac{\alpha}{4\pi}\right)^2\\
&{} - \biggl[ \frac{9}{2} L^3 + \frac{143}{6} L^2
+ \left(48 \pi^2 \log2 - 72 \zeta_3 - 55 \pi^2 + \frac{19363}{72}\right) L\\
&\hphantom{{}+\biggl[\biggr.}{}
+ 1024 a_4 + \frac{128}{3} \log^4 2
- 64 \pi^2 \log^2 2 - \frac{11792}{9} \pi^2 \log2
+ 20 \zeta_5\\
&\hphantom{{}+\biggl[\biggr.}{}
- 8 \pi^2 \zeta_3 + \frac{9494}{9} \zeta_3 + \frac{104}{45} \pi^4
+ \frac{259133}{405} \pi^2 + \frac{249887}{324}
\biggr]
\left(\frac{\alpha}{4\pi}\right)^3 + \cdots
\end{split}
\label{qed}
\end{equation}

In conclusion: we have derived the QCD/HQET matching coefficient
for the heavy-quark field with three-loop accuracy~(\ref{zmu}),
and the all-orders result in the large-$\beta_0$ limit.
The corresponding QED coefficient~(\ref{qed}) is gauge invariant.

\textbf{Acknowledgements.}
I am grateful to K.G.~Chetyrkin and M.~Steinhauser
for useful discussions and hospitality in Karlsruhe.
The work was supported by DFG through SFB/TR9.

\section*{Appendix. Gauge dependence of $Z_\psi$ and $\gamma_\psi$ in QED.}

The electron propagator $S(x)$ is related to the Landau-gauge propagator $S_L(x)$
($a_0=0$) by the formula~\cite{LK:56}
\begin{equation}
S(x) = S_L(x) e^{- i e_0^2 (\Delta(x)-\Delta(0))}\,,
\label{landau}
\end{equation}
where $\Delta(x)$ is the Fourier image of $\Delta(k)=a_0/(k^2)^2$,
and $\Delta(0)=0$ in dimensional regularization.
The electron field renormalization does not depend on its mass.
For simplicity, we shall consider the massless electron,
whose propagator has a single Dirac structure:
\begin{equation}
S(x) = S_0(x) e^{\sigma(x)}\,,
\label{Sx}
\end{equation}
where $S_0(x)$ is the $d$-dimensional free massless electron propagator.
Then
\begin{equation}
\sigma(x) = \sigma_L(x) + a_0 \frac{e_0^2}{(4\pi)^{d/2}}
\left(\frac{-x^2}{4}\right)^{\varepsilon}
\Gamma(-\varepsilon)\,,
\label{sigma}
\end{equation}
where the Landau-gauge $\sigma_L$ starts from $e_0^4$.
Re-expressing $\sigma$ via renormalized quantities we have
\begin{equation}
\sigma(x) = \sigma_L(x) + a(\mu) \frac{\alpha(\mu)}{4\pi}
\left(\frac{-\mu^2 x^2}{4}\right)^{\varepsilon}
e^{\gamma_E\varepsilon} \Gamma(-\varepsilon)\,.
\label{viaren}
\end{equation}
This must be equal to $\log Z_\psi+\sigma_r$,
where $\log Z_\psi(\alpha(\mu),a(\mu))$ contains only negative powers of $\varepsilon$,
and the renormalized $\sigma_r$ --- only non-negative.
Therefore,
\begin{equation}
\log Z_\psi(\alpha,a) = \log Z_L(\alpha) - a \frac{\alpha}{4\pi\varepsilon}\,.
\label{Zpsi}
\end{equation}
In QED $d\log(a(\mu)\alpha(\mu))/d\log\mu=-2\varepsilon$ exactly, and%
\footnote{I was informed by D.J.~Broadhurst and D.V.~Shirkov that this result
has been proved in some Russian article in the second half of 50s.
I am grateful to them for discussing this question;
unfortunately, I was unable to find this article.}
\begin{equation}
\gamma_\psi(\alpha,a) = 2 a \frac{\alpha}{4\pi} + \gamma_L(\alpha)\,,
\label{gammapsi}
\end{equation}
where the Landau-gauge $\gamma_L(\alpha)$ starts from $\alpha^2$.

\end{document}